\documentclass[prb,aps,twocolumn,superscriptaddress]{revtex4}
\usepackage{graphicx,amsmath,bm}
\usepackage{dcolumn}

\begin{document}

\title{Properties of Nambu-Goldstone Bosons \\ in a Single-Component Bose-Einstein Condensate}

\author{Takafumi Kita}
\affiliation{Department of Physics, Hokkaido University,
Sapporo 060-0810, Japan}
\date{\today}

\begin{abstract}
We theoretically study the properties of Nambu-Goldstone bosons in an interacting single-component Bose-Einstein condensate (BEC). We first point out that the proofs of Goldstone's theorem by Goldstone {\em et al}.\ [Phys.\ Rev.\ {\bf 127} (1962) 965] may be relevant to distinct massless modes of the BEC: whereas the first proof deals with the poles of the single-particle Green's function $\hat{G}$, the second one concerns those of the two-particle Green's function. Thus, there may be multiple Nambu-Goldstone bosons even in the single-component BEC with broken $U(1)$ symmetry. The second mode turns out to have an infinite lifetime in the long-wavelength limit in agreement with the conventional viewpoint. In contrast, the first mode from $\hat{G}$, i.e., the Bogoliubov mode in the weak-coupling regime, is shown to be a ``bubbling'' mode fluctuating temporally out of and back into the condensate. The substantial lifetime originates from an ``improper'' structure of the self-energy inherent in the BEC, which has been overlooked so far and will be elucidated here, and removes various infrared divergences pointed out previously.
\end{abstract}

\maketitle

\section{Introduction}

On the basis of the exact equations of motion for the single-particle Green's function $\hat{G}$ and condensate wave function,\cite{Kita09}
it was shown in a previous paper\cite{Kita10} that the poles of the two-particle Green's function $\underline{\cal K}$ in an interacting Bose-Einstein condensate (BEC) are generally not shared with those of $\hat{G}$. 
This implies that there may be multiple excitation branches even in a single-component BEC, contrary to the widely accepted viewpoint of attributing a unique phonon branch due to density fluctuations.\cite{Feynman54,GN64,SK74,WG74,Griffin93} 
Hence, it is desirable to clarify the basic properties of the poles of $\hat{G}$ and $\underline{\cal K}$. Here, we will perform it in terms of Goldstone's theorem\cite{Goldstone61,GSW62,GHK68,Brauner10} based on a self-consistent perturbation expansion that simultaneously satisfies Goldstone's theorem and conservation laws.\cite{Kita09}

Goldstone's theorem on spontaneously broken continuous symmetry has crucial importance over a wide range of fields in modern theoretical physics. According to a recent review,\cite{Brauner10} it is considered to predict a single massless mode called the Nambu-Goldstone boson for the single-component BEC with broken $U(1)$ symmetry. Since the mode is expected to dominate the thermodynamic properties of the BEC, clarifying the basic properties of the Nambu-Goldstone boson(s) forms an essential prerequisite to any theoretical investigations of the BEC. 
In this connection, it should be pointed out that basic proofs of Goldstone's theorem were given by Goldstone, Salam, and Weinberg (GSW) mainly in two different ways.\cite{GSW62} We will show that they may be relevant to distinct massless modes in general, contrary to the standard treatment\cite{GHK68,Brauner10} where no explicit distinction between the proofs seems to have been made.

The main results of the present paper are summarized as follows. For the single-component BEC, the two proofs by GSW \cite{GSW62} are relevant to the poles of different Green's functions. The first one deals with the single-particle Green's function and can be identified with the proof of the Hugenholtz-Pines theorem.\cite{HP59} The corresponding massless mode reduces to the Bogoliubov mode in the weak-coupling regime,\cite{Bogoliubov47} which has been regarded as a propagating mode with an infinite lifetime in the long-wavelength limit based on the work by Beliaev;\cite{Beliaev58b} see also refs.\ \onlinecite{NN78} and \onlinecite{PCCS97} on this point.
We will show, however, that it is essentially a ``bubbling'' mode with a short lifetime, i.e., a mode that fluctuates temporally out of and back into the condensate, originating from an ``improper'' structure of the self-energy inherent in the BEC. 
To put it another way, the first mode of the BEC is not a well-defined ``quasi-particle'' (``particle'') in the terminology of condensed matter (particle) physics. 
Both the improper structure of the self-energy and the resultant bubbling feature of the first mode have escaped consideration so far and will be studied in detail.
They also have the effect of removing infrared singularities pointed out by Gavoret and Nozi\`eres;\cite{GN64} 
hence $\underline{\cal K}$, for example,  can be calculated safely without any artificial procedures.
The second proof concerns a correlation function of three field operators.
Its poles will be shown (i) to be common with those of $\underline{\cal K}$ and (ii) to have a vanishing imaginary part in the long-wavelength limit. In other words, the second mode is an observable propagating one. Thus, the two modes in the BEC from the two proofs are completely different in character.

We now derive and elaborate on the above results. In \S 2, we recapitulate the main results of refs.\ \onlinecite{Kita09} and \onlinecite{Kita10} that are relevant to the present study. In \S 3, we consider the implications of Goldstone's theorem for the single-component BEC. Section 4 forms the core of the present study, in which we clarify the basic properties of the poles of $\hat{G}$ and $\underline{\cal K}$. Section 5 provides concluding remarks. We set $\hbar=k_{\rm B}=1$ throughout with $k_{\rm B}$ the Boltzmann constant.

\section{Summary of Previous Results}

The system we consider consists of identical particles with mass $m$ and spin $0$ described by the Hamiltonian
\begin{equation}
H=H_{0}+H_{\rm int},
\label{Hamil}
\end{equation}
with
\begin{subequations}
\label{Hamil2}
\begin{equation}
H_{0}=\int {\rm d}^{3}r_{1} \psi^{\dagger}({\bf r}_{1})K_{1}\psi({\bf r}_{1}),
\label{H_0}
\end{equation}
\begin{equation}
H_{\rm int}=\frac{1}{2}\int {\rm d}^{3}r\int {\rm d}^{3}r'\,\psi^{\dagger}({\bf r})
\psi^{\dagger}({\bf r}') V({\bf r}-{\bf r}')
 \psi({\bf r}')\psi({\bf r}) .
\label{H_int}
\end{equation}
\end{subequations}
Here $\psi^{\dagger}$ and $\psi$ are the creation and annihilation operators of the Bose field,  respectively, 
$K_{1}\equiv -{\hbar^{2}\nabla_{1}^{2}}/{2m}-\mu$
with $\mu$ the chemical potential,
and $V$ is the interaction potential. 
Though disregarded here, the effect of a trap potential may be included easily in $K_{1}$.
Let us introduce the Heisenberg representations of the field operators in the Matsubara formalism as  \cite{AGD63}
\begin{equation}
\left\{
\begin{array}{l}
\vspace{1mm}
\psi_{1}(1) \equiv e^{\tau_{1}H}\psi({\bf r}_{1})e^{-\tau_{1}H} \\
\psi_{2}(1) \equiv e^{\tau_{1}H}\psi^{\dagger}({\bf r}_{1})e^{-\tau_{1}H}
\end{array} \right. ,
\end{equation}
where the argument $1$ in the round brackets is defined by $1\equiv ({\bf r}_{1},\tau_{1})$, and the variable
$\tau_{1}$ lies in $0 \leq \tau_{1} \leq \beta\equiv 1/T$ with 
$T$ denoting the temperature.
The operators $\psi_{1}(1)$ and $\psi_{2}(1)$ were written as $\psi(1)$ and $\bar{\psi}(1)$ in ref.\ \onlinecite{Kita09}, respectively.
Distinguishing them by the subscript $i=1,2$ as above enables us to simplify the notation considerably, as seen below.

The operator $\psi_{i}(1)$ acquires a finite grand-canonical average in the condensed phase. Hence, it may be written as
\begin{equation}
\psi_{i}(1)=\Psi_{i}(1)+\phi_{i}(1),
\label{psi=Psi+psi-t}
\end{equation}
where $\Psi_{i}(1)$ is the condensate wave function
\begin{subequations}
\label{Psi-G}
\begin{equation}
\Psi_{i}(1)\equiv \frac{{\rm Tr}\,{\rm e}^{-\beta H}\psi_{i}(1)}{{\rm Tr}\,{\rm e}^{-\beta H}}\equiv \langle \psi_{i}(1)\rangle 
\label{Psi}
\end{equation}
with Tr the trace, and $\langle \phi_{i}(1)\rangle=0$ by definition. 
Note that $\Psi_{1}(1)=\Psi_{2}^{*}(1)=\Psi({\bf r}_{1})$ in equilibrium with $^{*}$ denoting the complex conjugate.
Using $\phi_{i}(1)$, we then introduce Green's functions as
\begin{equation}
G_{ij}(1,2)\equiv - \bigl< T_{\tau} \phi_{i}(1)\phi_{3-j}(2)\bigr>(-1)^{j-1} ,
\label{hatG}
\end{equation}
\end{subequations}
with $T_{\tau}$ the ``time''-ordering operator.\cite{AGD63}
They satisfy\cite{Kita09,Kita10}
\begin{eqnarray}
&&
\hspace{-10mm}
G_{ij}(1,2)= (-1)^{i+j-1}G_{3-j,3-i}(2,1)
\nonumber \\
&&\hspace{4mm}
=(-1)^{i+j}G_{ji}^{*}({\bf r}_{2}\tau_{1},{\bf r}_{1}\tau_{2}).
\label{G-symm}
\end{eqnarray}
The elements $G_{11}$ and $G_{12}$ were denoted as $G$ and $F$ in ref.\ \onlinecite{Kita09}, respectively, 
following the convention of superconductivity.\cite{AGD63}

We now summarize the main results of refs.\ \onlinecite{Kita09} and \onlinecite{Kita10} that are relevant to the present consideration.

\subsection{Hugenholtz-Pines relation and Dyson-Beliaev equation}

Let us introduce the $2\times 1$ vector $\vec{\Psi}\equiv[\Psi_{1}\,\Psi_{2}]^{\rm T}$
and $2\times 2$ Nambu matrix $\hat{G}\equiv (G_{ij})$,
where $^{\rm T}$ denotes transpose.
They obey\cite{Kita09} the generalized Gross-Pitaevskii equation (or Hugenholtz-Pines relation)
\begin{subequations}
\label{Dyson1}
\begin{equation}
\hat{G}^{-1}(1,\bar{2})\hat{\sigma}_{3}\vec{\Psi}(\bar{2})  =\vec{0} 
\label{HP}
\end{equation}
and the Dyson-Beliaev equation\cite{Beliaev58a,AGD63}
\begin{equation}
\hat{G}^{-1}(1,\bar{3})\hat{G}(\bar{3},2)
=\hat{\sigma}_{0}\delta(1,2) ,
\label{DB}
\end{equation}
\end{subequations}
respectively, where integrations over barred arguments are implied.
The quantity $\hat{G}^{-1}$ is defined by
\begin{equation}
\hat{G}^{-1}(1,2)\equiv \left(-\hat{\sigma}_{0}\frac{\partial}{\partial \tau_{1}}
-\hat{\sigma}_{3}K_{1}\right)\delta(1,2)
-\hat{\Sigma}(1,2) ,
\label{G^-1(1,2)}
\end{equation}
where $\hat{\sigma}_{0}$ and $\hat{\sigma}_{3}$ are the $2\times 2$ unit and third Pauli matrices, respectively, $K_{1}$ is given after eq.\ (\ref{Hamil2}), 
$\delta(1,2)\equiv \delta(\tau_{1}-\tau_{2})\delta({\bf r}_{1}-{\bf r}_{2})$, and $\hat{\Sigma}$ denotes the self-energy matrix.

Equation (\ref{HP}) was derived in Appendix A.4 of ref.\ \onlinecite{Kita09} by a formal argument using the global gauge transformation 
relevant to the broken $U(1)$ symmetry. In particular, the proof is free from a couple of assumptions by Hugenholtz and Pines\cite{HP59} that (i) the system is homogeneous and  (i) the self-energy is proper.
The first row of eq.\ (\ref{HP}) in equilibrium is written explicitly as 
$-K_{1}\Psi({\bf r}_{1})=\Sigma_{11}(1,\bar{2})\Psi(\bar{\bf r}_{2})-\Sigma_{12}(1,\bar{2})\Psi^{*}(\bar{\bf r}_{2})$.
By approximating $\Sigma_{11}(1,2)= 2g\delta(1,2)|\Psi({\bf r}_{1})|^{2}$ and $\Sigma_{12}(1,2)= g\delta(1,2)[\Psi({\bf r}_{1})]^{2}$ for $V({\bf r}_{1}-{\bf r}_{2})=g\delta({\bf r}_{1}-{\bf r}_{2})$, it reduces to the standard Gross-Pitaevskii equation.\cite{Gross61,Pitaevskii61,PS08}
Setting $\Psi\rightarrow\sqrt{n}_{0}$ and $K\rightarrow -\mu$ in the same equation with $n_{0}$ the condensate density, we also obtain the Hugenholtz-Pines relation for homogeneous systems.\cite{HP59,Kita09}

It was shown\cite{Kita09} that the elements of $\hat{\Sigma}$ are obtained from a single functional $\Phi$ by
\begin{subequations}
\label{Sigma-Phi}
\begin{equation}
\Sigma_{ij}(1,2)=-2\beta\frac{\delta\Phi}{\delta G_{ji}(2,1)},
\label{Sigma-Phi1}
\end{equation}
where $\Phi=\Phi[G,F,\bar{F},\Psi_{1},\Psi_{2}]$ with $G(1,2)=[G_{11}(1,2)-G_{22}(2,1)]/2$,
$F(1,2)=[G_{12}(1,2)+G_{12}(2,1)]/2$, and $\bar{F}(1,2)=-[G_{21}(1,2)+G_{21}(2,1)]/2$.
Here, we have incorporated the first symmetry of eq.\ (\ref{G-symm}) manifestly in $\Phi$;
this procedure is indispensable for discussing the two-particle irreducible vertices in terms of $\Phi$.
It follows from eq.\ (\ref{Sigma-Phi1}) that $\Sigma_{ij}$ also obey eq.\ (\ref{G-symm}) for $G_{ij}$.
There is another identity in terms of $\Phi$, which is given by eq.\ (21b) of ref.\ \onlinecite{Kita09}, i.e.,
\begin{equation}
\beta\frac{\delta \Phi}{\delta \Psi_{3-i}(1)}= \Sigma_{i\bar{j}}(1,\bar{2})(-1)^{\bar{j}-1}\Psi_{\bar{j}}(\bar{2}) .
\label{Sigma-Phi2}
\end{equation}
\end{subequations}
This is the relation that makes eq.\ (\ref{HP}) a stationarity condition of the thermodynamic potential 
$\Omega\equiv -\beta^{-1}\ln{\rm Tr}\,{\rm e}^{-\beta H}$ with respect to a variation of $\Psi_{i}$.\cite{Kita09}

For a given $\Phi$, eqs.\ (\ref{Dyson1}) and (\ref{Sigma-Phi1}) form a closed set of equations that can be used to determine $\vec{\Psi}$ and $\hat{G}$ self-consistently.

\subsection{Interaction energy}

By comparing eq.\ (\ref{Dyson1}) with the equations of motion for $\psi_{i}(1)$, 
it was shown that
the thermodynamic average of eq.\ (\ref{H_int}) can be expressed rigorously in terms of $\vec{\Psi}$, $\hat{G}$, and $\hat{\Sigma}$ as eq.\ (12) of Ref.\ \onlinecite{Kita09}.
In the present notation, it reads
\begin{equation}
\langle H_{\rm int}\rangle
=\frac{1}{4\beta}\Sigma_{\bar{i}\bar{j}}(\bar{1},\bar{2})
 \bigl[(-1)^{\bar{j}-1}\Psi_{\bar{j}}(\bar{2})\Psi_{3-\bar{i}}(\bar{1})-G_{\bar{j}\bar{i}}(\bar{2},\bar{1})\bigr] .
\label{<H_int>}
\end{equation}
In \S \ref{subsec:IP}, we will clarify a peculiar structure of the self-energy in the BEC based on this expression.

\subsection{Higher-order Green's functions}

We now consider a couple of higher-order Green's functions defined by
\begin{subequations}
\label{LK}
\begin{eqnarray}
&&\hspace{-10mm}
\langle 1_{i} |\underline{\tilde{\cal L}}|32_{kj}\rangle
 \equiv -\langle T_{\tau}\psi_{i}(1)\psi_{j}(2)\psi_{3-k}(3)\rangle (-1)^{k-1}
\nonumber \\
&&\hspace{11.5mm}
+ \Psi_{i}(1)\langle T_{\tau}\psi_{j}(2)\psi_{3-k}(3)\rangle (-1)^{k-1},
\label{L-def}
\end{eqnarray}
\begin{eqnarray}
&&\hspace{-6mm}
\langle 12_{ij} |\underline{\cal K}|43_{lk}\rangle
\equiv \langle T_{\tau}\psi_{i}(1)\psi_{3-j}(2)\psi_{k}(3)\psi_{3-l}(4)\rangle (-1)^{j+l}
\nonumber \\
&&\hspace{18.2mm}
-\langle T_{\tau}\psi_{i}(1)\psi_{3-j}(2)\rangle
 \langle T_{\tau}\psi_{k}(3)\psi_{3-l}(4)\rangle
\nonumber \\
&&\hspace{18.2mm}
\times (-1)^{j+l}.
\label{calK}
\end{eqnarray}
\end{subequations}
The two-particle Green's function $\underline{\cal K}$ can describe various collective modes such as density fluctuations.
To derive equations of motion for $\underline{\tilde{\cal L}}$ and $\underline{\cal K}$, let us add an extra perturbation given by the $S$ matrix
\begin{equation}
{\cal S}(\beta)\equiv T_{\tau}
\exp\!\left[-\frac{1}{2} \psi_{\,\bar{i}}(\bar{1})\psi_{3-\bar{j}}(\bar{2})(-1)^{\bar{j}-1} U_{\bar{j}\bar{i}}(\bar{2},\bar{1})\right]\! .
\label{calS}
\end{equation}
The condensate wave function and single-particle Green's function in the presence of the nonlocal potential $\hat{U}\equiv (U_{ij})$ are defined by \cite{AGD63}
\begin{subequations}
\label{Psi-calG^U}
\begin{equation}
\Psi_{i}^{U}(1)\equiv \frac{\langle T_{\tau}{\cal S}(\beta)\psi_{i}(1)\rangle}
{\langle{\cal S}(\beta)\rangle} ,
\label{Psi^U-def1}
\end{equation}
\begin{eqnarray}
&&\hspace{-12.5mm}
{\cal G}_{ij}^{U}(1,2)
\equiv -\frac{\langle T_{\tau}{\cal S}(\beta)\psi_{i}(1)\psi_{3-j}(2)\rangle (-1)^{j-1}}
{\langle{\cal S}(\beta)\rangle} 
\nonumber \\
&&\hspace{3mm}
= G_{ij}^{U}(1,2)-\Psi_{i}^{U}(1)\Psi_{3-j}^{U}(2)(-1)^{j-1} ,
\label{calG-def1}
\end{eqnarray}
\end{subequations}
where we have used the correspondent of eq.\ (\ref{psi=Psi+psi-t}) in the second equality of eq.\ (\ref{calG-def1}).
Quantities $\Psi_{i}^{U}(1)$ and $G_{ij}^{U}(1,2)$ reduce to eqs.\ (\ref{Psi}) and (\ref{hatG}) as $\hat{U}\rightarrow\hat{0}$, respectively.
It may be seen easily that $\underline{\tilde{\cal L}}$ of eq.\ (\ref{L-def}) is obtained from eq.\ (\ref{Psi^U-def1}) by
\begin{subequations}
\label{calLK-deriv}
\begin{equation}
\langle 1_{i} |\underline{\tilde{\cal L}}|32_{kj}\rangle= 2\frac{\delta\Psi_{i}^{U}(1)}{\delta U_{kj}(3,2)},
\label{calL-deriv}
\end{equation}
where the limit $\hat{U}\rightarrow\hat{0}$ is implied after the differentiation; we will use this convention below.
Similarly, eq.\ (\ref{calK}) is connected with eq.\ (\ref{calG-def1}) as
$\langle 12_{ij} |\underline{\cal K}|43_{lk}\rangle=
2{\delta {\cal G}_{ij}^{U}(1,2)}/{\delta U_{lk}(4,3)}$, i.e.,
\begin{eqnarray}
&&\hspace{-6mm}
\langle 12_{ij} |\underline{\cal K}|43_{lk}\rangle=  
2 \frac{\delta G_{ij}^{U}(1,2)}{\delta U_{lk}(4,3)}
-\bigl[\Psi_{i}(1)\langle 2_{3-j}|\underline{\tilde{\cal L}}|43_{lk}\rangle
\nonumber \\
&&\hspace{17.8mm}
+\langle 1_{i}|\underline{\tilde{\cal L}}|43_{lk}\rangle \Psi_{3-j}(2)\bigr] (-1)^{j-1} .
\label{calK-deriv}
\end{eqnarray}
\end{subequations}
Thus,  we only need to consider linear variations of
$\Psi_{i}^{U}$ and $G_{ij}^{U}$ to obtain $\underline{\tilde{\cal L}}$ and $\underline{\cal K}$ explicitly.

To carry this out, we start from eq.\ (\ref{Dyson1}). 
The perturbation given by eq.\ (\ref{calS}) produces in the right-hand side of eq.\ (\ref{G^-1(1,2)}) an extra term $-\hat{U}'(1,2)$ with\cite{MS59,BK61,Kita10} 
\begin{equation}
U_{ij}'(1,2)\equiv \frac{U_{ij}(1,2)+(-1)^{i+j-1}U_{3-j,3-i}(2,1)}{2} .
\label{U'}
\end{equation}
Varying $\hat{U}\rightarrow \hat{U}+\delta \hat{U}$ and subsequently setting $\hat{U}=\hat{0}$
in the resultant eq.\ (\ref{Dyson1}), we arrive at the first-order equations
\begin{subequations}
\label{Dyson2}
\begin{equation}
\hat{G}^{-1}(1,\bar{2})\hat{\sigma}_{3}\delta\vec{\Psi}^{U}(\bar{2})
=\bigl[\delta \hat{U}'(1,\bar{2})+\delta \hat{\Sigma}^{U}(1,\bar{2})\bigr]\hat{\sigma}_{3}\vec{\Psi}(\bar{2}),
\end{equation}
\begin{equation}
\hat{G}^{-1}(1,\bar{3})\delta\hat{G}^{U}(\bar{3},2)
=\bigl[\delta \hat{U}'(1,\bar{3}) +\delta \hat{\Sigma}^{U}(1,\bar{3})\bigr] \hat{G}(\bar{3},2).
\end{equation}
\end{subequations}
We next express the first-order self-energy $\delta \hat{\Sigma}^{U}$ in terms of $\delta \hat{G}^{U}$ and $\delta \vec{\Psi}^{U}$ as
\begin{eqnarray}
\delta \Sigma_{ij}^{U}(1,2)\!\!\!&=&\!\!\! -2\langle 12_{ij}|\underline{\Gamma}^{(4)}|\bar{4}\bar{3}_{\bar{l}\bar{k}}\rangle\delta G_{\bar{l}\bar{k}}^{U}(\bar{4},\bar{3})
\nonumber \\
\!\! & &\!\!\! 
+2\langle 12_{ij}|\underline{\Gamma}^{(3)}|\bar{3}_{\bar{k}}\rangle (-1)^{\bar{k}-1}\delta \Psi_{\bar{k}}^{U}(\bar{3}),
\label{dSigma}
\end{eqnarray}
with
\begin{subequations}
\label{Gamma}
\begin{equation}
\langle 12_{ij}|\underline{\Gamma}^{(4)}|43_{lk}\rangle\!\equiv\! -\frac{1}{2}\frac{\delta\Sigma_{ij}(1,2)}{\delta G_{lk}(4,3)}
\!=\! \beta\frac{\delta^{2}\Phi }{\delta G_{ji}(2,1)\delta G_{lk}(4,3)},
\label{Gamma^(4)}
\end{equation}
\begin{eqnarray}
&&\hspace{-12mm}
\langle 12_{ij}|\underline{\Gamma}^{(3)}|3_{k}\rangle \equiv \frac{1}{2}\frac{\delta\Sigma_{ij}(1,2)}{\delta \Psi_{k}(3)}(-1)^{k-1}
\nonumber \\
&&\hspace{9mm}
=2(-1)^{k+\bar{l}} \langle 12_{ij}|\underline{\Gamma}^{(4)}|\bar{4}3_{\bar{l},3-k}\rangle
\Psi_{\bar{l}}(\bar{4}) ,
\label{Gamma^(3)}
\end{eqnarray}
where eq.\ (\ref{Sigma-Phi}) has been used to derive the second expressions.
We also introduce a set of matrices
\begin{eqnarray}
\hspace{-5mm}\langle 1_{i}|\underline{\tilde{\Gamma}}^{(3)}|32_{kj}\rangle
\!&\equiv&\! 2(-1)^{\bar{l}-1} \langle 1\bar{4}_{i\bar{l}}|\underline{\Gamma}^{(4)}|32_{kj}\rangle
\Psi_{\bar{l}}(\bar{4}) 
\nonumber \\
\!&=&\!(-1)^{i}\langle 23_{jk}|\underline{\Gamma}^{(3)}|1_{3-i}\rangle,
\label{Gamma^(3)t}
\\
\hspace{-5mm}\langle 1_{i}|\hat{\Gamma}^{(2)}|2_{j}\rangle\!&\equiv&\!  2(-1)^{\bar{k}-1}
\langle 1\bar{3}_{i\bar{k}}|\underline{\Gamma}^{(3)}|2_{j}\rangle\Psi_{\bar{k}}(\bar{3}),
\label{Gamma^(2)}
\\
\hspace{-5mm}\langle 12_{ij}|\underline{\chi}^{(0)}|43_{lk}\rangle\!&\equiv&\!
G_{il}(1,4)G_{kj}(3,2)+(-1)^{k+l-1}
\nonumber \\
& &\! \times G_{i,3-k}(1,3)G_{3-l,j}(4,2),
\label{chi^(0)}
\\
\hspace{-5mm}\langle 12_{ij}|\underline{\Psi}^{(3)}|3_{k}\rangle\!&\equiv&\!  
(-1)^{j+k}\bigl[ \Psi_{i}(1) \delta_{k,3-j}\delta(3,2)
\nonumber \\
& &\!
+ \delta_{ki}\delta(3,1)\Psi_{3-j}(2)\bigr],
\label{uPsi}
\\
\hspace{-5mm}\langle 1_{i}|\underline{\tilde{\Psi}}^{(3)}|32_{kj}\rangle\!&\equiv&\!  
(-1)^{j-1}\bigl[ \delta_{ik}\delta(1,3)\Psi_{j}(2)
\nonumber \\
& &\!
+ \delta_{i,3-j}\delta(1,2)\Psi_{3-k}(3)\bigr] ,
\label{uPsit}
\end{eqnarray}
\end{subequations}
and vectors
\begin{equation}
\langle 12_{ij}|\delta\vec{G}^{U}\equiv \delta G_{ij}^{U}(1,2),
\hspace{5mm}
\langle 12_{ij}|\delta\vec{U}\equiv \delta U_{ij}(1,2) .
\label{Vec3}
\end{equation}
The quantities in eqs.\ (\ref{Gamma^(4)})-(\ref{Gamma^(2)}) form ``irreducible'' vertices of our BEC, as seen below in eq.\ (\ref{BS}).
Let us substitute eqs.\ (\ref{U'}) and (\ref{dSigma}) into eq.\ (\ref{Dyson2}) and
make use of eqs.\ (\ref{Gamma^(3)t}) and (\ref{Gamma^(2)}) as well as the symmetry
$\langle 12_{ij}|\underline{\Gamma}^{(4)}|43_{lk}\rangle =\langle 34_{kl}|\underline{\Gamma}^{(4)}|21_{ji}\rangle =
(-1)^{i+j-1}\langle 21_{3-j,3-i}|\underline{\Gamma}^{(4)}|43_{lk}\rangle$ originating from eqs.\ (\ref{G-symm}) and  (\ref{Gamma^(4)}).
We thereby obtain a closed set of equations for $\delta \vec{\Psi}^{U}$ and $\delta\vec{G}^{U}$ as\cite{Kita10}
\begin{eqnarray*}
\hat{\sigma}_{3}\delta \vec{\Psi}^{U}=\frac{1}{2}\hat{G}\underline{\tilde{\Psi}}^{(3)}\delta\vec{U}
+\hat{G}\hat{\Gamma}^{(2)}\hat{\sigma}_{3}\delta \vec{\Psi}^{U}
-\hat{G}\underline{\tilde{\Gamma}}^{(3)}\delta \vec{G}^{U},
\\
\delta \vec{G}^{U}=\frac{1}{2}\underline{\chi}^{(0)}\delta\vec{U}
+\underline{\chi}^{(0)}\underline{\Gamma}^{(3)}\hat{\sigma}_{3}\delta \vec{\Psi}^{U}
-\underline{\chi}^{(0)}\underline{\Gamma}^{(4)}\delta \vec{G}^{U}.
\end{eqnarray*}
By considering eq.\ (\ref{calLK-deriv}), they can be expressed equivalently in terms of
$\underline{\tilde{\cal L}}$ and $\underline{{\cal K}}'\equiv 2\delta\vec{G}^{U}\!/\delta\vec{U}$
as
\begin{equation}
\begin{bmatrix}
\vspace{2mm}
\hat{\sigma}_{3}\underline{\tilde{\cal L}}\,\, \\
\underline{\cal K}'
\end{bmatrix}= 
\begin{bmatrix}
\vspace{1mm}
\hat{G}\underline{\tilde{\Psi}}^{(3)}\\
\underline{\chi}^{(0)}
\end{bmatrix}
+ 
\begin{bmatrix}
\vspace{1mm}
\hat{G} & \underline{\tilde{0}}\\
\underline{0} & \underline{\chi}^{(0)}
\end{bmatrix}
\begin{bmatrix}
\vspace{1mm}
\hat{\Gamma}^{(2)} & -\underline{\tilde{\Gamma}}^{(3)}\\
\underline{\Gamma}^{(3)} & -\underline{\Gamma}^{(4)}
\end{bmatrix}
\begin{bmatrix}
\vspace{2mm}
\hat{\sigma}_{3}\underline{\tilde{\cal L}}\,\, \\
\underline{\cal K}'
\end{bmatrix} ,
\label{BS}
\end{equation}
which may be regarded as the Bethe-Salpeter equation\cite{SB51} for the condensed Bose system and
can be solved formally with respect to $\underline{\tilde{\cal L}}$ and $\underline{\cal K}'$.\cite{Kita10}
Using eqs.\ (\ref{calK-deriv}) and (\ref{uPsi}), we finally arrive at the explicit expressions\cite{Kita10}
\begin{subequations}
\label{LK-solution}
\begin{eqnarray}
\underline{\tilde{\cal L}} 
\!\!\!&\equiv&\!\!\!\hat{\sigma}_{3}\hat{\chi}^{({\rm c})}
\bigl(\underline{\tilde{\Psi}}^{(3)}-\underline{\tilde{\Gamma}}^{(3)}\underline{\chi}^{(4)}\bigr),
\label{uL}
\\
\underline{\cal K}\!\!\!&\equiv&\!\!\! \underline{\chi}^{({\rm q})}
+\underline{\chi}^{(4)}\underline{\Gamma}^{(3)}\hat{\chi}^{({\rm c})}\underline{\tilde{\Psi}}^{(3)}
+\underline{\Psi}^{(3)}\hat{\chi}^{({\rm c})}
\underline{\tilde{\Gamma}}^{(3)}\underline{\chi}^{(4)}
\nonumber \\
& &\!\!\!
-\underline{\Psi}^{(3)}\hat{\chi}^{({\rm c})}\underline{\tilde{\Psi}}^{(3)} ,
\label{ucalK}
\end{eqnarray}
\end{subequations}
where $\underline{\chi}^{(4)}$, $\hat{\chi}^{({\rm c})}$, and $\underline{\chi}^{({\rm q})}$ are defined by
\begin{subequations}
\label{chicq}
\begin{eqnarray}
\underline{\chi}^{(4)}\!\!\!&\equiv&\!\!\!\bigl(\underline{\chi}^{(0)-1}+\underline{\Gamma}^{(4)}\bigr)^{-1} ,
\label{chi^(4)}
\\
\hat{\chi}^{({\rm c})}\!\!\!&\equiv&\!\!\! \left(\hat{G}^{-1}-\hat{\Gamma}^{(2)}+
\underline{\tilde{\Gamma}}^{(3)}\underline{\chi}^{(4)}
\underline{\Gamma}^{(3)}\right)^{\!\!-1},
\label{chic}
\\
 \underline{\chi}^{({\rm q})}\!\!\!&\equiv&\!\!\!\left[\underline{\chi}^{(4)-1}+
\underline{\Gamma}^{(3)}\left(\hat{G}^{-1}-\hat{\Gamma}^{(2)}\right)^{\!\!-1}
\underline{\tilde{\Gamma}}^{(3)}\right]^{-1},
\label{chiq}
\end{eqnarray}
\end{subequations}
with superscripts $^{\rm c}$ and $^{\rm q}$ denoting ``condensate'' and ``quasi-particle,'' respectively.
It is worth pointing out that
the poles of $\underline{\chi}^{(4)}$ in eq.\ (\ref{LK-solution}) are cancelled by those of $\underline{\chi}^{(4)}$ in the denominator of eq.\ (\ref{chic}).

We observe from eq.\ (\ref{LK-solution}) that the function $\hat{\chi}^{({\rm c})}$, which is characteristic of the BEC, appears in both $\underline{\tilde{\cal L}}$ and $\underline{\cal K}$.
This $\hat{\chi}^{({\rm c})}$ has been identified as $\hat{G}$ in the diagrammatic analyses 
of $\underline{\cal K}$ for homogeneous systems.\cite{GN64,SK74,WG74,Griffin93}
However, $\hat{\chi}^{({\rm c})}$ of eq.\ (\ref{chic}) is clearly different from $\hat{G}$ 
due to the additional contribution $-\hat{\Gamma}^{(2)}+\underline{\tilde{\Gamma}}^{(3)}\underline{\chi}^{(4)}\underline{\Gamma}^{(3)}$  in the denominator.
The properties of the poles of $\hat{G}$ and $\hat{\chi}^{({\rm c})}$, especially their differences,  will be one of the main topics below.

\subsection{The functional $\Phi$\label{subsec:Phi}}

The key quantity in the present formalism is the functional $\Phi=\Phi[G,F,\bar{F},\Psi_{1},\Psi_{2}]$ with 
\begin{subequations}
\label{GF-def}
\begin{eqnarray}
G(1,2)\!&\equiv&\!\bigl[{G_{11}(1,2)-G_{22}(2,1)}\bigr]/{2},
\\
F(1,2)\!&\equiv&\!\bigl[{G_{12}(1,2)+G_{12}(2,1)}\bigr]/{2},
\\
\bar{F}(1,2)\!&\equiv&\!-\bigl[{G_{21}(1,2)+G_{21}(2,1)}\bigr]/{2} .
\end{eqnarray}
\end{subequations}
Indeed, the self-energies and irreducible vertices are determined from this $\Phi$ by eqs.\ (\ref{Sigma-Phi1}) and (\ref{Gamma^(4)})-(\ref{Gamma^(2)}), respectively;
using them, we can obtain $(\vec{\Psi},\hat{G})$ and $(\underline{\tilde{\cal L}},\underline{\cal K})$ self-consistently by
eqs.\ (\ref{Dyson1}) and (\ref{BS}), respectively. 
More generally, choosing an approximate $\Phi$ amounts to fixing the whole Bogoliubov-Born-Green-Kirkwood-Yvon (BBGKY) hierarchy.\cite{Cercignani88,Kita10s}

\begin{figure}[t]
        \begin{center}
                \includegraphics[width=0.85\linewidth]{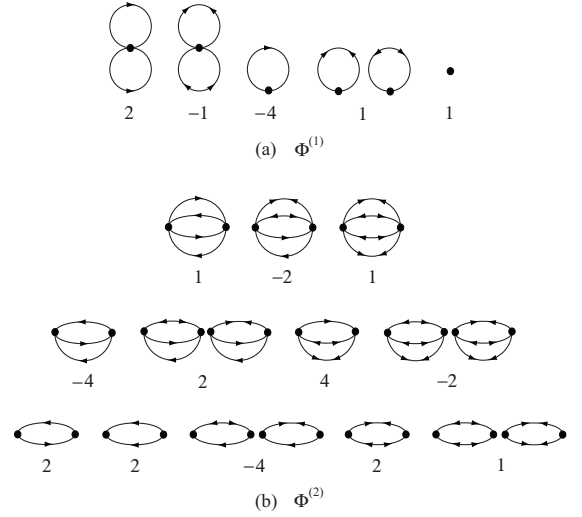}
        \end{center}
        \vspace{-4mm}
        \caption{Diagrammatic representations of (a) $\Phi^{(1)}$ and (b) $\Phi^{(2)}$. A filled circle denotes $\Gamma^{(0)}$, and a line with an arrow (two arrows) represents $G$
         (either $F$ or $\bar{F}$) in eq.\ (\ref{GF-def}) as in the theory of superconductivity.\cite{AGD63} 
         The number below each diagram indicates its relative weight.\label{Fig1}}
\end{figure}

It was shown\cite{Kita09} that $\Phi$ can be constructed order by order perturbatively in terms of the symmetrized
bare vertex
\begin{eqnarray}
&&\hspace{-10mm}
\Gamma^{(0)}(11',22')
\equiv V({\bf r}_{1}-{\bf r}_{2})\delta(\tau_{1}-\tau_{2})
\nonumber \\
&&\hspace{14mm}
\times[\delta(1,1')\delta(2,2')+\delta(1,2')\delta(2,1')],
\label{Gamma^(0)}
\end{eqnarray}
so as to satisfy eqs.\ (\ref{Sigma-Phi}) and (\ref{<H_int>}) simultaneously.
The first- and second-order contributions are given diagrammatically in Figs.\ \ref{Fig1}(a) and \ref{Fig1}(b), respectively.
The corresponding analytic expressions read
\begin{subequations}
\label{Phi}
\begin{eqnarray}
&&\hspace{-6mm}
\Phi^{(1)}=\frac{1}{4\beta} \Gamma^{(0)}(\bar{1}\bar{1}',\bar{2}\bar{2}') \bigl[ 2G(\bar{1},\bar{1}')G(\bar{2},\bar{2}')
\nonumber \\
&&\hspace{6mm}
-F(\bar{1},\bar{2})\bar{F}(\bar{1}',\bar{2}')
-4G(\bar{1},\bar{1}')\Psi_{1}(\bar{2})\Psi_{2}(\bar{2}')
\nonumber \\
&&\hspace{6mm}
+F(\bar{1},\bar{2})\Psi_{2}(\bar{1}')\Psi_{2}(\bar{2}')
+\bar{F}(\bar{1}',\bar{2}')\Psi_{1}(\bar{1})\Psi_{1}(\bar{2})
\nonumber \\
&&\hspace{6mm}
+\Psi_{2}(\bar{1}')\Psi_{2}(\bar{2}')\Psi_{1}(\bar{2})\Psi_{1}(\bar{1}) \bigr] ,
\label{Phi^(1)}
\end{eqnarray}
\begin{eqnarray}
&&\hspace{-2mm}
\Phi^{(2)}
=-\frac{1}{8\beta}
\Gamma^{(0)}(\bar{1}\bar{1}',\bar{2}\bar{2}')\Gamma^{(0)}(\bar{3}\bar{3}',\bar{4}\bar{4}')
\nonumber \\
&&\hspace{9mm}
\times
\bigl\{G(\bar{2},\bar{3}')G(\bar{3},\bar{2}')G(\bar{1},\bar{4}')G(\bar{4},\bar{1}')
\nonumber \\
&&\hspace{9mm}
-2\bar{F}(\bar{2}',\bar{3}')F(\bar{2},\bar{3})G(\bar{1},\bar{4}')G(\bar{4},\bar{1}')
\nonumber \\
&&\hspace{9mm}
+\bar{F}(\bar{2}',\bar{3}')F(\bar{2},\bar{3})\bar{F}(\bar{1}',\bar{4}')F(\bar{1},\bar{4})
\nonumber \\
&&\hspace{9mm}
-4G(\bar{2},\bar{3}')G(\bar{3},\bar{2}')G(\bar{1},\bar{4}')\Psi_{1}(\bar{4})\Psi_{2}(\bar{1}')
\nonumber \\
&&\hspace{9mm}
+4\bar{F}(\bar{2}',\bar{3}')F(\bar{2},\bar{3})G(\bar{1},\bar{4}')\Psi_{1}(\bar{4})\Psi_{2}(\bar{1}')
\nonumber \\
&&\hspace{9mm}
+2\bigl[\bar{F}(\bar{2}',\bar{3}')\Psi_{1}(\bar{2})\Psi_{1}(\bar{3})
+F(\bar{2},\bar{3})\Psi_{2}(\bar{2}')\Psi_{2}(\bar{3}')\bigr]
\nonumber \\
&&\hspace{9mm}
\times
G(\bar{1},\bar{4}')G(\bar{4},\bar{1}')
\nonumber \\
&&\hspace{9mm}
-2\bigl[\bar{F}(\bar{2}',\bar{3}')\Psi_{1}(\bar{2})\Psi_{1}(\bar{3})+
F(\bar{2},\bar{3})\Psi_{2}(\bar{2}')\Psi_{2}(\bar{3}')\bigr]
\nonumber \\
&&\hspace{9mm}
\times
\bar{F}(\bar{1}',\bar{4}')F(\bar{1},\bar{4})
\nonumber \\
&&\hspace{9mm}
+2G(\bar{2},\bar{3}')G(\bar{3},\bar{2}')\Psi_{1}(\bar{1})\Psi_{2}(\bar{4}')\Psi_{1}(\bar{4})\Psi_{2}(\bar{1}')
\nonumber \\
&&\hspace{9mm}
+2G(\bar{2},\bar{3}')\Psi_{1}(\bar{3})\Psi_{2}(\bar{2}')G(\bar{1},\bar{4}')\Psi_{1}(\bar{4})\Psi_{2}(\bar{1}')
\nonumber \\
&&\hspace{9mm}
-4\bigl[\bar{F}(\bar{2}',\bar{3}')\Psi_{1}(\bar{2})\Psi_{1}(\bar{3})
+F(\bar{2},\bar{3})\Psi_{2}(\bar{2}')\Psi_{2}(\bar{3}')\bigr]
\nonumber \\
&&\hspace{9mm}
\times
G(\bar{1},\bar{4}')\Psi_{1}(\bar{4})\Psi_{2}(\bar{1}')
\nonumber \\
&&\hspace{9mm}
+F(\bar{2},\bar{3})\Psi_{2}(\bar{3}')\Psi_{2}(\bar{2}')F(\bar{1},\bar{4})
\Psi_{2}(\bar{4}')\Psi_{2}(\bar{1}')
\nonumber \\
&&\hspace{9mm}
+\bar{F}(\bar{2}',\bar{3}')\Psi_{1}(\bar{3})\Psi_{1}(\bar{2})\bar{F}(\bar{1}',\bar{4}')
\Psi_{1}(\bar{4})\Psi_{1}(\bar{1})
\nonumber \\
&&\hspace{9mm}
+2\bar{F}(\bar{2}',\bar{3}')F(\bar{2},\bar{3})\Psi_{1}(\bar{1})\Psi_{2}(\bar{4}')\Psi_{1}(\bar{4})\Psi_{2}(\bar{1}')
\bigr\} .
\nonumber \\
\label{Phi^(2)}
\end{eqnarray}
\end{subequations}

\section{Goldstone's Theorem for Single-Component BEC}

We now discuss Goldstone's theorem\cite{Goldstone61,GSW62,GHK68,Brauner10}
in terms of the single-component BEC.
Bose-Einstein condensation is a textbook example of broken $U(1)$ symmetry
described by the Mexican hat potential\cite{PS95}
as well as a prototype of the off-diagonal long-range order.\cite{PO56,Yang62} 
These facts may make the single-component BEC one of the most desirable grounds for studying the implications of Goldstone's theorem. 
It will be shown that the basic proofs of Goldstone's theorem by GSW \cite{GSW62} are relevant to the poles of distinct Green's functions
for the BEC.

The first proof concerns the single-particle Green's function $\hat{G}$ defined by eq.\ (\ref{hatG}), which obeys the Dyson-Beliaev equation (\ref{DB});
our $G_{ij}$ corresponds to the complete propagator $\Delta_{ij}'$ of GSW.\cite{GSW62}
Now, the conclusion of the first proof can be stated in the present context as follows:
the operator $\hat{G}^{-1}$ in eq.\ (\ref{DB}) should also determine the condensate wave function 
$\vec{\Psi}=[\Psi_{1}\,\Psi_{2}]^{\rm T}$ by eq.\ (\ref{HP}).
Equation (\ref{HP}), i.e., the Hugenholtz-Pines relation for general inhomogeneous systems,\cite{Kita09} may be identified with eq.\ (18) of GSW,
which manifestly tells us that poles of $\hat{G}$ contain a ``massless'' mode.
It should be noted, however, that eq.\ (\ref{Dyson1}) does not say anything about the nature of the mode, which will be elucidated in \S \ref{sec:NG}.

We next focus on the second proof by GSW involving the use of a commutation relation.\cite{GSW62} Although it is applicable only to homogeneous systems, it has been referred to more frequently as the standard proof of Goldstone's theorem.\cite{GHK68,Brauner10}
In the context of the BEC, the proof is relevant to
\begin{equation}
\rho_{n\psi}({\bf p},\omega)\equiv  \int {\rm d}^{3}r  \int_{-\infty}^{\infty}{\rm d}t\,{\rm e}^{-i{\bf p}\cdot{\bf r}+i\omega t}\langle [n({\bf r},t),\psi({\bf 0})] \rangle ,
\label{rho}
\end{equation}
where $n({\bf r},t)\equiv {\rm e}^{iHt}\psi^{\dagger}({\bf r})\psi({\bf r}){\rm e}^{-iHt}$ is the density operator in the Heisenberg representation, and we have chosen the argument of $\psi$ arbitrarily at the coordinate origin based on the translational symmetry.  Setting ${\bf p}={\bf 0}$ in eq.\ (\ref{rho}) yields 
$$\rho_{n\psi}({\bf 0},\omega)=\int_{-\infty}^{\infty}{\rm d}t\, {\rm e}^{i\omega t}\langle[{\rm e}^{iHt}N{\rm e}^{-iHt},\psi({\bf 0})] \rangle,$$ 
where 
$N\equiv \int {\rm d}^{3}r\, \psi^\dagger({\bf r})\psi({\bf r})$
is the particle-number operator, i.e.,  the generator of the gauge transformation relevant to the broken $U(1)$ symmetry. Since $[N,H]=0$ and $[N,\psi({\bf 0})]=-\psi({\bf 0})$, we immediately obtain
\begin{equation}
\rho_{n\psi}({\bf 0},\omega)=-2\pi\sqrt{n_{0}}\,\delta(\omega) ,
\label{rho(0)}
\end{equation}
which is finite for $\sqrt{n_{0}}\equiv \langle \psi_{i}\rangle\neq 0$ with a sharp peak at $\omega=0$.
Assuming that $\rho_{n\psi}({\bf p},\omega)$ is continuous for $({\bf p},\omega)\rightarrow ({\bf 0},0)$, which is justified here,\cite{GHK68,Brauner10}
we conclude that $\rho_{n\psi}({\bf p},\omega)$ contains a massless mode with an infinite lifetime in the long-wavelength limit. 

It is clear from eq.\ (\ref{rho}) that the second proof deals with a correlation function of three field operators, whose poles may be different from those of eq.\ (\ref{hatG}). 
This is confirmed by relating $\rho_{n\psi}$ to $\underline{\tilde{\cal L}}$ of eq.\ (\ref{L-def}).
Consider the Fourier transform
\begin{equation}
\tilde{\cal L}_{1,11}({\bf p},z_{\ell})\equiv \int {\rm d}^{3}r_{21}\int_{0}^{\beta}{\rm d}\tau_{21}  \langle 1_{1}|\underline{\tilde{\cal L}}|22_{11}\rangle
{\rm e}^{-i{\bf p}\cdot{\bf r}_{21}+z_{\ell}\tau_{21}},
\label{L(p)}
\end{equation}
where $\tau_{21}\equiv \tau_{2}-\tau_{1}$, ${\bf r}_{21}\equiv {\bf r}_{2}-{\bf r}_{1}$, and  $z_{\ell}\equiv 2\ell\pi i/\beta$ with $\ell= 0,\pm 1,\cdots$.
Using the Lehmann representation,\cite{AGD63} one can show that
eq.\ (\ref{rho}) is connected with eq.\ (\ref{L(p)}) as 
\begin{equation}
\rho_{n\psi}({\bf p},\omega)=-2{\rm Im}\tilde{\cal L}_{1,11}({\bf p},\omega+i0_{+}),
\end{equation}
where $0_{+}$ denotes an infinitesimal positive constant.
On the other hand, $\underline{\tilde{\cal L}}$ can be expressed as eq.\ (\ref{uL}), whose poles are shared with those of $\underline{\cal K}$ in eq.\ (\ref{ucalK})
through $\hat{\chi}^{({\rm c})}$.
Hence, the second mode also belongs to the two-particle Green's function $\underline{\cal K}$.
Finally, $\hat{\chi}^{({\rm c})}$ in eq.\ (\ref{chic}) is not $\hat{G}$ itself due to the extra contribution 
$-\hat{\Gamma}^{(2)}+\underline{\tilde{\Gamma}}^{(3)}\underline{\chi}^{(4)}\underline{\Gamma}^{(3)}$ in the denominator.

Thus, the two proofs by GSW may generally predict two distinct ``massless'' modes for the single-component BEC. As already mentioned, this fact has apparently been overlooked so far;\cite{GHK68,Brauner10} it is certainly not incorporated in the counting of Nambu-Goldstone bosons, which is based on the second proof.\cite{Brauner10}

\begin{figure}[t]
        \begin{center}
                \includegraphics[width=0.85\linewidth]{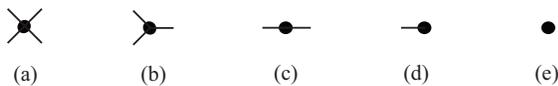}
        \end{center}
        \vspace{-4mm}
        \caption{Diagrammatic representation of the two-body interaction in terms of the excitation field $\phi_{i}$. A line and a filled circle denote $\phi_{i}$ and $\Gamma^{(0)}$, respectively. \label{Fig2}}
\end{figure}
\begin{figure}[t]
        \begin{center}
                \includegraphics[width=0.8\linewidth]{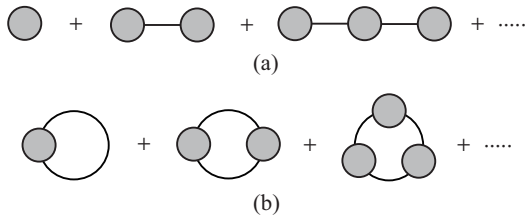}
        \end{center}
        \vspace{-4mm}
        \caption{Diagrammatic structures of $\langle H_{\rm int}\rangle$. A shaded circle denotes the proper contribution, which cannot be separated into two parts by cutting a single line.\label{Fig3}}
\end{figure}

\section{Properties of Nambu-Goldstone Bosons\label{sec:NG}}

Having clarified the differences between the two proofs of Goldstone's theorem by GSW,\cite{GSW62} we now study the properties of the corresponding two massless modes. 

\subsection{Improper structure of self-energy in BEC\label{subsec:IP}}

We start by clarifying a peculiar structure of the self-energy inherent in the BEC. Substituting eq.\ (\ref{psi=Psi+psi-t}) into eq.\ (\ref{H_int}), we can express
$H_{\rm int}$ diagrammatically as shown in Fig.\ \ref{Fig2} in terms of the excitation field $\phi_{i}$ and the symmetrized vertex $\Gamma^{(0)}$ of eq.\ (\ref{Gamma^(0)}). Diagrams (b)-(e) are characteristic of the BEC. In particular, (b)-(d) give rise to the structures shown in Fig.\ \ref{Fig3} in the thermodynamic average $\langle H_{\rm int}\rangle$. 
On the other hand, $\langle H_{\rm int}\rangle$ can be expressed analytically as eq.\ (\ref{<H_int>}), which was derived rigorously on the basis of eq.\ (\ref{Dyson1}).\cite{Kita09} Comparing Fig.\ \ref{Fig3} with eq.\ (\ref{<H_int>}), we are led to the conclusion that the self-energy itself has the structure of Fig.\ \ref{Fig3}(a), i.e., the structure called ``improper'' in the normal state, \cite{LW60}  as it can be separated into two parts by cutting a single line. 

Improper contributions to $\hat{\Sigma}$ have been excluded in the classic studies of the BEC,\cite{HP59,GN64,SK74,WG74,Griffin93,Beliaev58b,NN78,PCCS97,AGD63} 
where the self-energy appears to have been introduced by using the concept ``proper''  \cite{LW60} without asking explicitly 
whether the relevant quantity is identical with the one in the Dyson-Beliaev equation. 
However, the above consideration has clarified that the distinction between ``proper'' and ``improper'' cannot be used to construct the self-energy defined by the Dyson-Beliaev equation (\ref{DB}).

\begin{figure}[t]
        \begin{center}
                \includegraphics[width=0.85\linewidth]{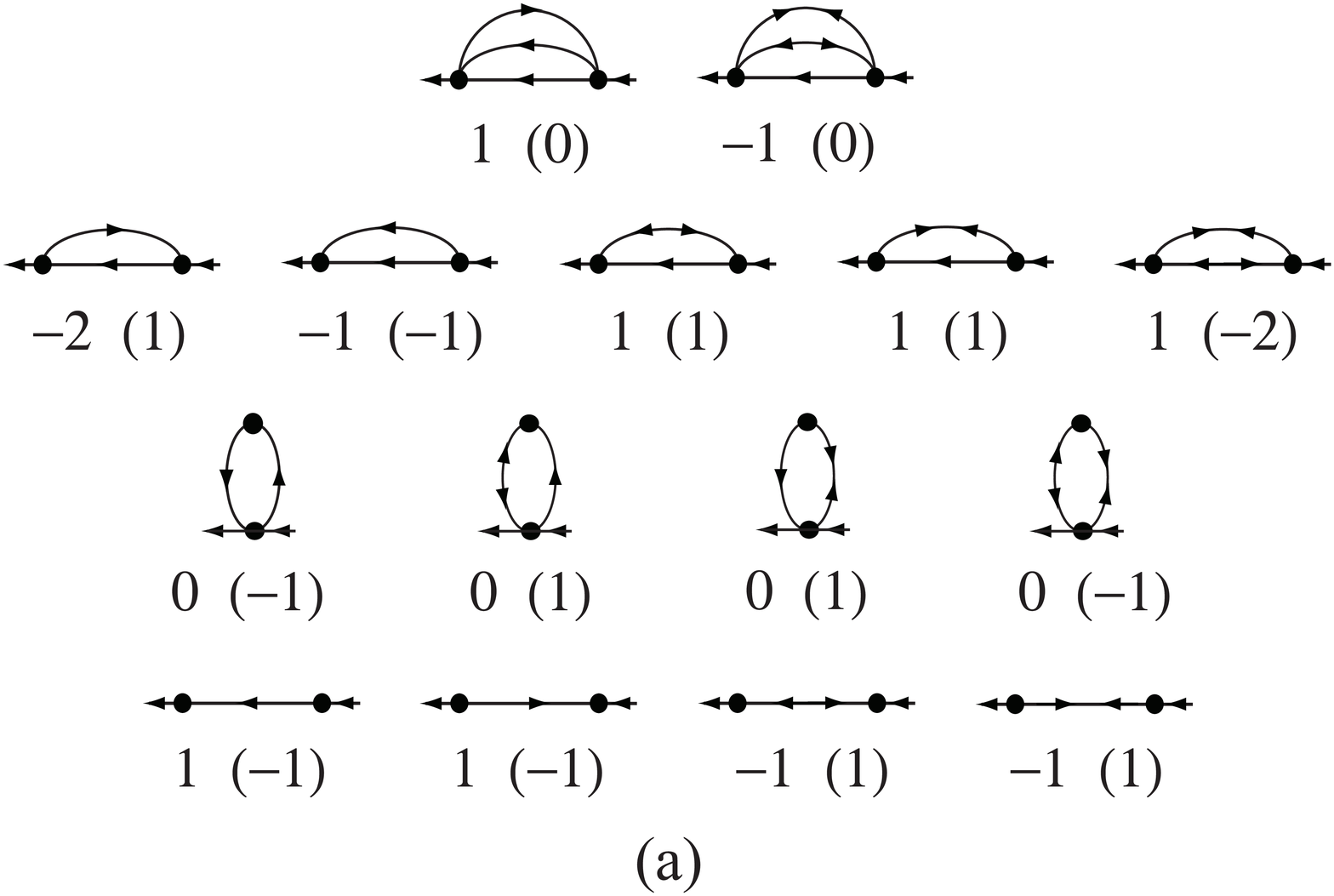}
        \end{center}
        \begin{center}
                \includegraphics[width=0.85\linewidth]{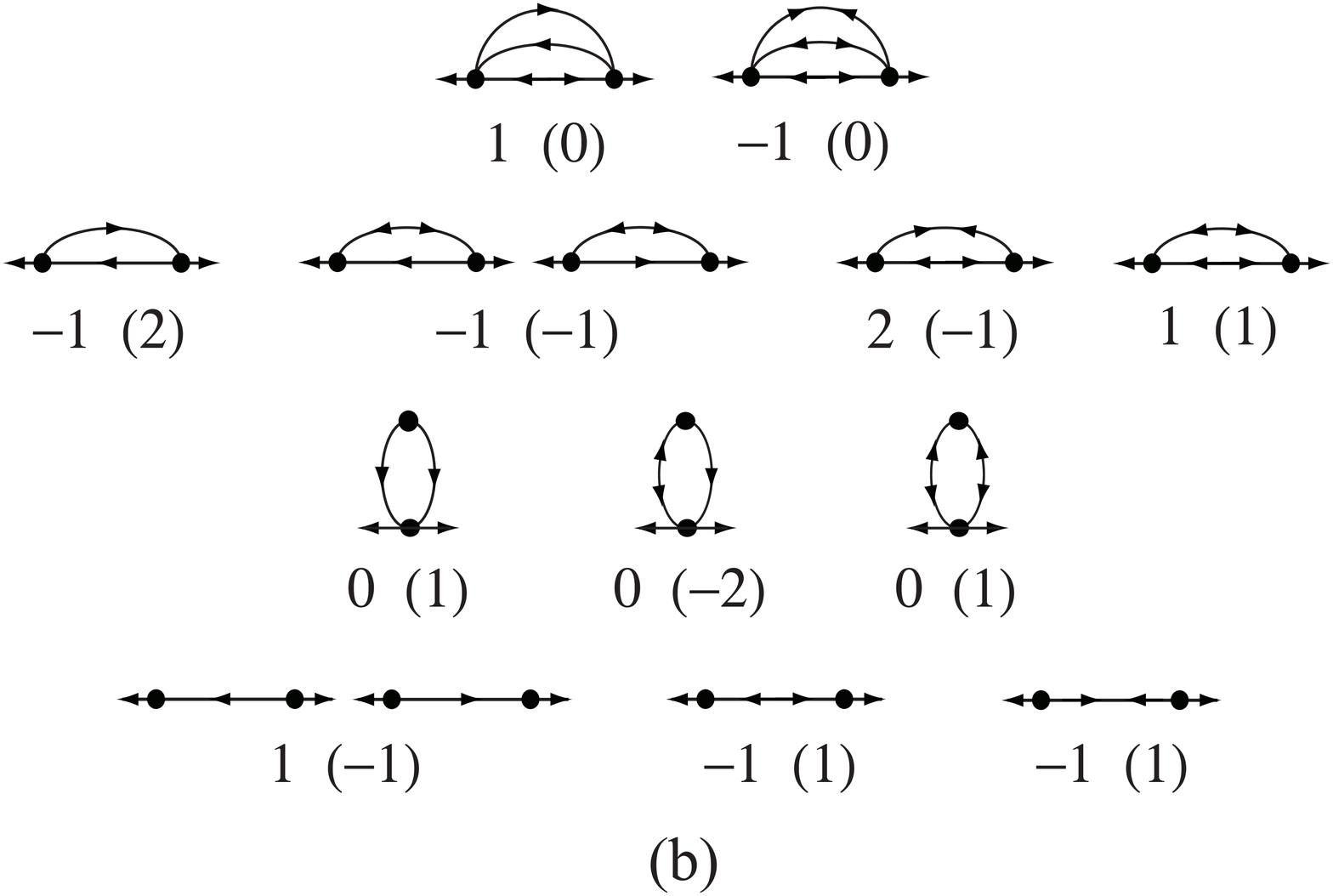}
        \end{center}
        \vspace{-4mm}
\caption{Second-order diagrams for: (a) $\Sigma_{11}$ and $\Gamma^{(2)}_{11}$; (b) $\Sigma_{12}$ and $\Gamma^{(2)}_{12}$.
A line with an arrow (two arrows) represents $G$ (either $F$ or $\bar{F}$) in eq.\ (\ref{GF-def}) as in  the theory of superconductivity.\cite{AGD63}
The left number (the right number in round brackets) below each diagram indicates its relative weight in $\Sigma_{1j}$ ($\Gamma^{(2)}_{1j}$). Multiplying each number by $1/2$ yields the absolute weight. \label{Fig4}}
\end{figure}
\begin{figure}[t]
        \begin{center}
                \includegraphics[width=0.95\linewidth]{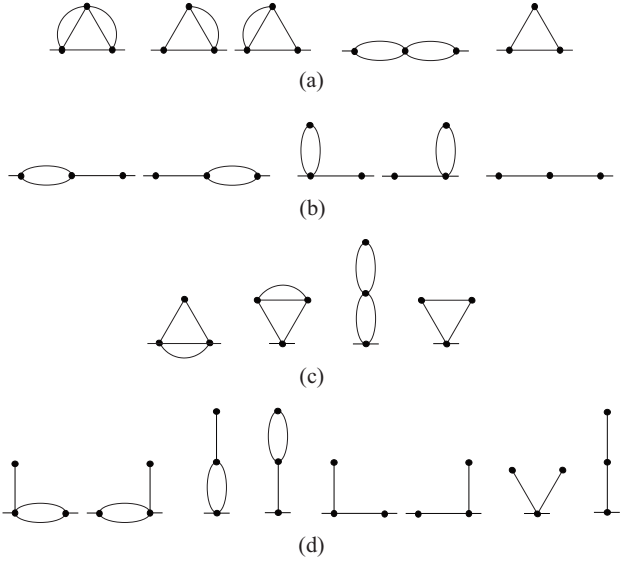}
        \end{center}
        \vspace{-4mm}
        \caption{Various third-order diagrams for $\hat{\Sigma}$ and $\hat{\Gamma}^{(2)}$. (a) Diagrams that contribute to both $\hat{\Sigma}$ and $\hat{\Sigma}+\hat{\Gamma}^{(2)}$.
        (b) Improper diagrams for $\hat{\Sigma}$ that disappear from $\hat{\Sigma}+\hat{\Gamma}^{(2)}$.
        (c) Diagrams peculiar to $\hat{\Gamma}^{(2)}$. (d) Diagrams for $\hat{\Gamma}^{(2)}$ with null total contribution.
        \label{Fig5}}
\end{figure}

The necessity of improper contributions to $\hat{\Sigma}$ was also pointed out recently on the basis of a self-consistent perturbation expansion simultaneously satisfying Goldstone's theorem and conservation laws.\cite{Kita09}
According to this expansion, the self-energies may be obtained from a single functional $\Phi$ by eq.\ (\ref{Sigma-Phi1}), and the
functional $\Phi$ can be constructed order by order uniquely so as to satisfy eqs.\ (\ref{Sigma-Phi}) and (\ref{<H_int>}) simultaneously.
Equations (\ref{Phi^(1)}) and (\ref{Phi^(2)}) present explicit expressions of the first- and second-order contributions, respectively.
This functional $\Phi$ also determines ``irreducible'' vertices in the Bethe-Salpeter equation (\ref{BS}) as eqs.\ (\ref{Gamma^(4)})-(\ref{Gamma^(2)}).
In particular, eq.\ (\ref{Gamma^(2)}) can be written in terms of $\Phi$ as 
\begin{eqnarray}
\Gamma^{(2)}_{ij}(1,2)\!&=&\! 4\beta\frac{\delta^{2}\Phi}{\delta G_{\bar{k}i}(\bar{3},1)\delta G_{\bar{l},3-j}(\bar{4},2)}
\Psi_{\bar{k}}(\bar{3})\Psi_{\bar{l}}(\bar{4})
\nonumber \\
& &\! \times(-1)^{j-1+\bar{k}+\bar{l}}.
\label{Gamma^(2)-2}
\end{eqnarray}
This differentiation may also be carried out graphically by removing from the diagrams of $\Phi$ a couple of lines with the second subscripts $i$ and $3-j$ in all possible ways.
Figure \ref{Fig4} enumerates the second-order diagrams for $\Sigma_{1j}$ and $\Gamma^{(2)}_{1j}$ ($j=1,2$) obtained from eq.\ (\ref{Phi^(2)}) or Fig.\ \ref{Fig1}(b) by eqs.\ (\ref{Sigma-Phi1}) and (\ref{Gamma^(2)-2}), respectively. Among them, the diagrams in the bottom lines of Figs.\ \ref{Fig4}(a) and \ref{Fig4}(b) have the structure called ``improper'' in the normal state. \cite{LW60} Using the number below each diagram to indicate its relative weight, however, one can easily check that they cancel out exactly in the sum $\hat{\Sigma}+\hat{\Gamma}^{(2)}$. Thus, they do not contribute to the denominator of eq.\ (\ref{chic}). The presence of improper diagrams in $\hat{\Sigma}$ and their disappearance from $\hat{\Sigma}+\hat{\Gamma}^{(2)}$ are expected to be a general feature of the BEC. We have also confirmed the statements in the third order of the self-consistent perturbation expansion using eqs.\ (B1) and (B5) of Ref.\ \onlinecite{Kita09} and eqs.\ (\ref{Sigma-Phi1}) and (\ref{Gamma^(2)-2}) above; the results are 
summarized briefly in Fig.\ \ref{Fig5} without arrows.

\subsection{Spectral function of a dilute Bose gas}

The improper structure of the self-energy $\hat{\Sigma}$ inherent in the BEC has profound implications for the nature of its first mode determined as the poles of $\hat{G}$. This mode, i.e., the Bogoliubov mode in the weak-coupling regime, \cite{Bogoliubov47} has been regarded as a propagating mode with an infinite lifetime in the long-wavelength limit. \cite{Beliaev58b,AGD63,PCCS97} However, the improper structure gives rise to considerable broadening of the poles of $\hat{G}$, changing the nature of the massless mode dramatically, as seen below. We call this mode, i.e., a mode with a substantial lifetime due to improper contributions to the self-energy, a ``bubbling'' mode based on the fact that it fluctuates temporally out of and back into the condensate reservoir.

To convince ourselves of this statement, consider a homogeneous system with the contact interaction
\begin{equation}
V({\bf r}_{1}-{\bf r}_{2})=g\delta({\bf r}_{1}-{\bf r}_{2}),
\label{V-contact}
\end{equation}
and expand Green's function as
\begin{equation}
\hat{G}(1,2)=\frac{1}{\beta{\cal V}}\sum_{\vec{p}}
\hat{G}(\vec{p})\,{\rm e}^{i{\bf p}\cdot({\bf r}_{1}-{\bf r}_{2})-z_{\ell}(\tau_{1}-\tau_{2})} ,
\end{equation}
where ${\cal V}$ is the volume and $\vec{p}\equiv ({\bf p},z_{\ell})$ with $z_{\ell}\equiv 2\ell\pi i/\beta$ ($\ell= 0,\pm 1,\cdots$).
Noting that $\Gamma^{(0)}\rightarrow 2g$ and $\Psi_{i}\rightarrow \sqrt{n_{0}}$ for the homogeneous system with
$n_{0}$ denoting the condensate density, 
we can express the improper diagrams in the bottom lines of Figs.\ \ref{Fig4}(a) and \ref{Fig4}(b)
in $\vec{p}$ space as ($j=1,2$)
\begin{equation}
\Sigma_{1j}^{(2{\rm ip})}(\vec{p})=2g^{2}n_{0}^{2}\bigl[G_{11}({\vec{p}})-G_{22}({\vec{p}})-G_{12}({\vec{p}})+G_{21}({\vec{p}})
\bigr].
\label{Sigma^(2)-ip}
\end{equation}
Now, if the retarded Green's function 
$\hat{G}^{\rm R}({\bf p},\omega)\equiv\hat{G}({\bf p}, \omega+i0_{+})$
had a pole at $\omega=E_{\bf p}$, it would in turn bring about an infinite imaginary part at $\omega=E_{\bf p}$ in the retarded self-energies obtained from eq.\ (\ref{Sigma^(2)-ip}) by $z_{\ell}\rightarrow  \omega+i0_{+}$. Thus, a sharp peak in the spectral function
\begin{equation}
A_{\bf p}(\omega)\equiv 
-2{\rm Im}G_{11}({\bf p},\omega+i0_{+})
\label{A_p(w)}
\end{equation}
is impossible in the presence of improper contributions to $\hat{\Sigma}$. 
In other words, no well-defined quasi-particle can exist in the single-particle channel. 

To see this more clearly, let us introduce an approximation for $\hat{G}(\vec{p})$ where we incorporate only eq.\ (\ref{Sigma^(2)-ip}) from the second-order along with the first-order self-energy $\hat{\Sigma}^{(1)}$ studied in detail in Ref.\ \onlinecite{Kita05}. This approximation may be quantitatively excellent for a dilute Bose gas; it will certainly enable us to qualitatively and explicitly see how improper contributions affect the excitation spectrum. 
Although eq.\ (\ref{Sigma^(2)-ip}) is apparently second-order in $g$, 
the self-consistency condition on $\Sigma_{1j}^{(2{\rm ip})}$ will be seen to yield a first-order contribution, changing the massless Bogoliubov spectrum dramatically. In contrast, other proper diagrams of the second order will only quantitatively modify the first-order Bogoliubov spectrum to a tiny extent for the dilute Bose gas. Hence, they may be neglected for the present purpose. 

In general, eq.\ (\ref{DB}) can be inverted for homogeneous systems to give
\begin{equation}
\hat{G}({\vec{p}})
=\frac{1}{D_{\vec{p}}}\!
\begin{bmatrix}
\vspace{1mm}
z_{\ell}+\varepsilon_{{\bf p}}+\bar{\Sigma}_{\vec{p}}-\mu & \Delta_{\vec{p}}
\\
-\bar{\Delta}_{\vec{p}} & z_{\ell}-\varepsilon_{\bf p}-\Sigma_{\vec{p}}+\mu 
\end{bmatrix},
\label{hatG_p}
\end{equation}
where $\varepsilon_{{\bf p}}\equiv {\bf p}^{2}/2m$,  ${\Sigma}_{\vec{p}}\equiv\Sigma_{11}(\vec{p})$, ${\Delta}_{\vec{p}}\equiv\Sigma_{12}(\vec{p})$, $\bar{\Sigma}_{\vec{p}}\equiv -\Sigma_{22}(\vec{p})=\Sigma_{-\vec{p}^{*}}^{*}$, $\bar{\Delta}_{\vec{p}}\equiv -\Sigma_{21}(\vec{p})=\Delta_{-\vec{p}^{*}}^{*}$,
and $D_{\vec{p}}$ denotes the determinant
\begin{equation}
D_{\vec{p}}=(z_{\ell}+\varepsilon_{{\bf p}}+\bar{\Sigma}_{\vec{p}}-\mu)(z_{\ell}-\varepsilon_{\bf p}-\Sigma_{\vec{p}}+\mu)
+ \Delta_{\vec{p}}\bar{\Delta}_{\vec{p}} .
\label{D_p}
\end{equation}
The first-order self-energies $\Sigma^{(1)}$ and $\Delta^{(1)}$ have no $\vec{p}$ dependence for the contact potential of eq.\ (\ref{V-contact}).
Their expressions are obtained by performing the differentiation of eq.\ (\ref{Sigma-Phi1}) with eq.\ (\ref{Phi^(1)})
and carrying out the Fourier transform as
\begin{equation}
\Sigma^{(1)}=2gn, \hspace{5mm}
\Delta^{(1)}=g\left[n_{0}-\frac{1}{\beta{\cal V}}\sum_{\vec{p}}G_{12}(\vec{p})\right],
\label{Delta^(1)}
\end{equation}
where
\begin{equation}
n\equiv n_{0}-\frac{1}{\beta{\cal V}}\sum_{\vec{p}}G_{11}(\vec{p})
\label{n}
\end{equation}
denotes the particle density.
On the other hand, the elements of eq.\ (\ref{Sigma^(2)-ip}) satisfy
\begin{equation}
\Sigma_{1j}^{(2{\rm ip})}(\vec{p})=
\bar{\Sigma}_{1j}^{(2{\rm ip})}(\vec{p})\equiv\Delta^{(2)}_{\vec{p}}.
\label{Delta^(2)-def}
\end{equation}
Adopting the approximation mentioned above, 
we can express the self-energies in eqs.\ (\ref{hatG_p}) and (\ref{D_p}) as 
\begin{subequations}
\label{eq-2nd}
\begin{equation}
 \Sigma_{\vec{p}}=\bar{\Sigma}_{\vec{p}}=\Sigma^{(1)}+\Delta_{\vec{p}}^{(2)} , \hspace{5mm}
 \Delta_{\vec{p}}=\bar{\Delta}_{\vec{p}}=\Delta^{(1)}+\Delta_{\vec{p}}^{(2)} .
\label{Sigma-Delta(p)}
\end{equation}
Hence, eq.\ (\ref{HP}) reduces to 
\begin{equation}
\mu=\Sigma^{(1)}-\Delta^{(1)}.
\label{HP-2nd}
\end{equation}
\end{subequations}
Using these results in eq.\ (\ref{hatG_p}), we obtain a simplified expression for $\hat{G}(\vec{p})$ with 
$\bar{\Sigma}_{\vec{p}}-\mu={\Sigma}_{\vec{p}}-\mu=\bar{\Delta}_{\vec{p}}=\Delta_{\vec{p}}$.
Let us substitute it into eq.\ (\ref{Sigma^(2)-ip}) and use eq.\ (\ref{Delta^(2)-def}). 
We then find that eq.\ (\ref{Sigma^(2)-ip}) forms a quadratic equation for $\Delta^{(2)}_{\vec{p}}$ as
\begin{equation}
\bigl(\Delta^{(2)}_{\vec{p}}\bigr)^{2}-\bigl(D^{(1)}_{\vec{p}}/2\varepsilon_{{\bf p}}\bigr)\Delta^{(2)}_{\vec{p}}+2g^{2}n_{0}^{2}=0
\label{Delta^(2)-eq}
\end{equation}
with
\begin{equation}
D^{(1)}_{\vec{p}}\equiv \bigl(z_{\ell}-E_{{\bf p}}^{{\rm B}}\bigr)\bigl(z_{\ell}+E_{{\bf p}}^{{\rm B}}\bigr),
\end{equation}
where
\begin{equation}
E_{{\bf p}}^{{\rm B}}\equiv \sqrt{\varepsilon_{{\bf p}}(\varepsilon_{{\bf p}}+2\Delta^{(1)})}
\label{E_p^B}
\end{equation}
is the Bogoliubov spectrum\cite{AGD63,Griffin93,PS08,Kita05} with a linear dispersion for $|{\bf p}|\rightarrow 0$.
By imposing $\Delta^{(2)}_{\vec{p}} \rightarrow 0$ as $|z_{\ell}|\rightarrow \infty$, \cite{AGD63} eq.\ (\ref{Delta^(2)-eq}) is solved easily as
\begin{eqnarray}
\Delta^{(2)}_{\vec p}=D^{(1)}_{\vec{p}}/4\varepsilon_{\bf p}
- \!\left[\bigl(D^{(1)}_{\vec{p}}/4\varepsilon_{\bf p}\bigr)^{2}-2g^{2}n_{0}^{2}\,\right]^{1/2}.
\end{eqnarray}
Finally, the spectral function is obtained by eq.\ (\ref{A_p(w)}), i.e., 
\begin{equation}
A_{\bf p}(\omega)=-2{\rm Im}\frac{z_{\ell}+\varepsilon_{{\bf p}}+\Delta_{\vec{p}}}{z_{\ell}^{2}-\varepsilon_{{\bf p}}(\varepsilon_{{\bf p}}+2\Delta_{\vec{p}})}
\biggr|_{z_{\ell}\rightarrow \omega+i0_+},
\label{A_p(w)-2}
\end{equation}
which is a function of $(\omega,\varepsilon_{\bf p},\Delta^{(1)},gn_0)$.

It is worth pointing out that the spectral function (\ref{A_p(w)}) can be negative for condensed Bose systems.
This may be realized by expressing it in the Lehmann representation\cite{AGD63}
\begin{eqnarray}
&&\hspace{-10mm}
A_{\bf p}(\omega)=2\pi \sum_{\nu\nu'}\,{\rm e}^{\beta(\Omega-{\cal E}_{\nu})}\!\left[1-{\rm e}^{-\beta ({\cal E}_{\nu'}-{\cal E}_{\nu})}\right]
\nonumber \\
&& \hspace{4mm}\times
\langle \varphi_{\nu} |c_{\bf p}|\varphi_{\nu'} \rangle \langle\varphi_{\nu'} |c_{\bf p}^{\dagger}
|\varphi_{\nu} \rangle \delta(\omega-{\cal E}_{\nu'}+{\cal E}_{\nu}),
\label{A-Lehmann}
\end{eqnarray}
where ${\cal E}_{\nu}$ and $\varphi_{\nu}$ are an eigenvalue of the Hamiltonian given by eq.\ (\ref{Hamil}) and its eigenfunction, 
$\Omega\equiv -\beta^{-1}\ln\sum_{\nu}{\rm e}^{-\beta {\cal E}_{\nu}}$ is the thermodynamic potential, and $c_{\bf p}$ and $c_{\bf p}^{\dagger}$ denote the annihilation and creation operators
of the plane-wave state ${\bf p}$, respectively. One may expect from eq.\ (\ref{A-Lehmann}) that ${\cal E}_{\nu'}-{\cal E}_{\nu}\geq 0$ holds for any finite matrix element  $\langle \varphi_{\nu'}|c_{\bf p}^{\dagger}
|\varphi_{\nu} \rangle$ so that $A_{\bf p}(\omega)\geq 0$. 
However, such an argument is no longer valid for condensed Bose systems where $c_{\bf p}^{\dagger}$ is expressed as a linear combination of quasi-particle creation and annihilation operators.\cite{AGD63,Kita09}
A classic counterexample is given by Bogoliubov theory, whose spectral function is obtained 
 by setting $\Delta_{\vec{p}}^{(2)}\rightarrow 0$ in eq.\ (\ref{A_p(w)-2}) as
\begin{equation}
A_{{\bf p}}^{{\rm B}}(\omega)=\pi [(\alpha_{\bf p}+1)\delta(\omega-E_{{\bf p}}^{{\rm B}})-(\alpha_{\bf p}-1)\delta(\omega+E_{{\bf p}}^{{\rm B}})],
\label{A^B}
\end{equation}
with $\alpha_{\bf p}\equiv (\varepsilon_{\bf p}+\Delta^{(1)})/E_{{\bf p}}^{{\rm B}}>1$.
This function consists of a couple of sharp quasi-particle peaks at $\omega=\pm E_{\bf p}^{\rm B}$, and the one at $\omega=- E_{\bf p}^{\rm B}$ is indeed negative.
Note also that eq.\ (\ref{A^B}) satisfies the general sum rule $\int_{-\infty}^{\infty}A_{{\bf p}}(\omega){\rm d}\omega =2\pi$ exactly, as it should.

 \begin{figure}[t]
        \begin{center}
                \includegraphics[width=0.85\linewidth]{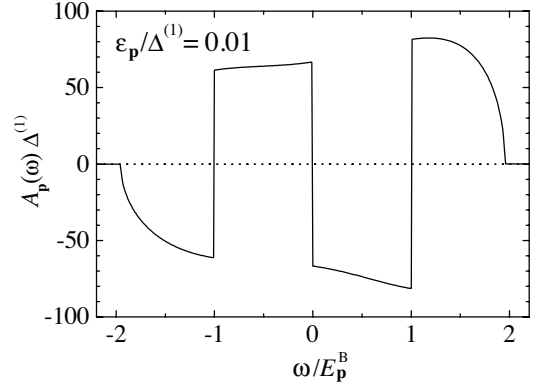}
        \end{center}
        \vspace{-4mm}
        \caption{Normalized spectral function $A_{\bf p}(\omega)\Delta^{(1)}$ for $\varepsilon_{\bf p}/\Delta^{(1)}=0.01$ as a function of $\omega/E_{{\bf p}}^{{\rm B}}$. \label{Fig6}}
\end{figure}

We have calculated eq.\ (\ref{A_p(w)-2}) by replacing $gn_0\rightarrow \Delta^{(1)}$ in the formula, which is justified for the dilute Bose gas near $T=0$ as seen from eq.\ (\ref{Delta^(1)}). 
The resultant dimensionless spectral function $A_{\bf p}(\omega)\Delta^{(1)}$ becomes a function of only $\omega/\Delta^{(1)}$ and $\varepsilon_{\bf p}/\Delta^{(1)}$. 
This approximation for $A_{\bf p}(\omega)$ will be excellent for weakly interacting Bose gases of $gn /T_0 \ll 1$ and $\Delta^{(1)}\propto gn$ at low temperatures, where $T_0=3.31 n^{2/3}/m$ denotes the transition temperature of the ideal Bose gas.\cite{Kita05}

Figure \ref{Fig6} shows $A_{\bf p}(\omega)$ as a function of $\omega$ for $\varepsilon_{\bf p} /\Delta^{(1)} =0.01$, which lies in the region of the linear Bogoliubov dispersion in eq.\ (\ref{E_p^B}).
Comparing Fig.\ \ref{Fig6} with eq.\ (\ref{A^B}),
we clearly observe that the couple of sharp quasi-particle peaks at $\omega/E_{{\bf p}}^{{\rm B}}=\pm 1$ in Bogoliubov theory are substantially blurred.
Thus, the first mode is no longer a well-defined quasi-particle after incorporating the lowest-order improper contributions to the self-energy.
Although the present consideration is based on the second-order analysis,
the substantial lifetime is expected to be an essential consequence of the improper contribution, as may be justified by the general argument after eq.\ (\ref{Sigma^(2)-ip}) or checked by investigating higher-order terms.
Since the particle density $n$ is given by eq.\ (\ref{n}), we may identify this ``bubbling'' mode as the temporal number-phase fluctuations of spontaneously broken $U(1)$ symmetry. The absence of sharp poles in $\hat{G}(\vec{p})$ also has the effect of removing infrared divergences pointed out by Gavoret and Nozi\`eres. \cite{GN64} Hence, the two-particle Green's function, for example, can be calculated safely without any artificial procedures by incorporating improper terms into the self-energy.

Concerning the second mode, the improper contributions disappear from the denominator of eq.\ (\ref{chic}), as already mentioned after eq.\ (\ref{Gamma^(2)-2}).
Thus, $\hat{\chi}^{({\rm c})}$ will have a sharp quasi-particle peak in agreement with eq.\ (\ref{rho(0)}), which may be observed as a propagating mode with density fluctuations. 
We emphasize once again that this mode should be distinguished clearly from the Bogoliubov mode.\cite{Bogoliubov47}

\section{Concluding Remarks}

Perhaps it is not unreasonable to generally expect that single-particle and collective excitations of a many-particle system, determined as the poles of its single-particle and two-particle Green's functions, respectively, are different in character. 
Indeed, this is exactly the case for Fermi systems where the two excitations even obey different statistics of the Fermi and Bose types, respectively, and it is the single-particle excitation that is responsible for most of their thermodynamic properties.\cite{AGD63} 

The present paper has shown that the above statement will also hold true for the single-component BEC, contrary to the conventional viewpoint of attributing a unique phonon branch of collective density fluctuations as its excitation.\cite{Feynman54,GN64,SK74,WG74,Griffin93} 
A characteristic of the system is that the two modes obey the same Bose statistics and may contribute equally to various phenomena. Thus, there remains a fundamental question: which mode dominates the thermodynamic, transport, and other properties of the single-component BEC? The answer may depend crucially on the interaction strength for the respective phenomena. In this context, it is worth pointing out that single-particle excitations are completely neglected in the variational wave function of Feynman for superfluid $^4$He,\cite{Feynman54} which incorporates only the collective density fluctuations as may be justified in the strong-coupling region. However, single-particle excitations should also be relevant in the weak-coupling region as in the case of the ideal Bose gas. Further investigations are required to clarify this point.

\begin{acknowledgments}
This work is supported by a Grant-in-Aid for Scientific Research 
from the Ministry of Education, Culture, Sports, Science and Technology
of Japan.
\end{acknowledgments}

\end{document}